\documentclass[aps,prx,twocolumn]{revtex4-2}
\usepackage{amsmath}
\usepackage{amssymb}
\usepackage{xcolor}
\usepackage{graphicx}
\usepackage{bm}
\usepackage{bbm}
\usepackage{dcolumn}
\usepackage{graphics}
\usepackage{array}
\usepackage{float}
\usepackage{tikz}
\usepackage{hyperref}
\usepackage{physics}

\begin{document}
\title{A Mobile Impurity in the Kitaev Chain: Phase Diagram and Signatures of Topology}
\author{A.V.~Sadovnikov$^{1,2}$, M.S.~Bahovadinov$^{1,3}$, A.N.~Rubtsov$^{1,4}$ and A.A.~Markov$^{1}$}
\affiliation{$^{1}$ Russian Quantum Center, Skolkovo IC, Moscow 121205, Russia} 
\affiliation{$^2$ Moscow Institute of Physics and Technology, 117303 Moscow, Russia}
\affiliation{$^3$ Condensed-Matter Physics Laboratory, HSE University, 101000 Moscow, Russia}
\affiliation{$^4$ Department of Physics, Lomonosov Moscow State University, 119991 Moscow, Russia}

\begin{abstract}

We study the physics of a mobile impurity immersed in a $1d$ topological superconductor. We discuss the system's phase diagram obtained with exact diagonalization. We argue that the character of the transition from a weak to strong coupling regime depends on the phase of the host superconductor. A smooth crossover between a weakly coupled polaron and a molecular state is observed in the topological phase. In contrast, the impurity undergoes a sharp phase transition in a topologically trivial background.
\end{abstract}

\maketitle

\section{Introduction}

Non-trivial topological state of a gapped many-body system
manifests itself in observables of quite distinct physical nature. Perhaps most notably, the topology is reflected in bulk transport properties \cite{klitzing1980new}, in the presence of gapless boundary modes \cite{halperin1982quantized} and in the non-trivial statistics of excitations \cite{arovas1984fractional}. Each of these ways to probe the topology may fail an observer or be unavailable, depending on a topological class and physical setting. Thus, bulk topological invariants are not easily measurable for some classes of topological insulators, for example protected by spatial symmetries \cite{fu2011topological}. Bulk topological invariants connected to the transport observables are not easily accessible in cold atomic experiments. Zero energy boundary modes can have a non-topological nature \cite{yu2021non}. Such problems are most well-known and controversial in one-dimensional topological superconductors \cite{kitaev2001unpaired}.

Kitaev chain model realisation in semiconductor nanowires turned out to be a much more complicated problem than it had been expected. Yet, more challenging task nowadays is to convince the community that such a realization has been performed successfully \cite{frolov2021quantum}. Bulk topological invariant of the Kitaev chain is not that straightforwardly connected to a measurable observable \cite{kitaev2001unpaired,akhmerov2011quantized}. Therefore most of the experimental effort has been directed at demonstrating the appearance of topologically protected Majorana zero modes at the edges of a wire. Numerous claims of the Majorana zero modes observation have been made \cite{mourik2012signatures, gul2018ballistic, zhang2018retracted,nadj2014observation} and consequently cast into doubt \cite{lee2012zero,sau2015bound,frolov2021quantum, frolov2024comment,hess2022prevalence}. Several directions to overcome the difficulties have been proposed \cite{yazdani2023hunting}. Naturally, one might aim at strengthening the evidence of the topological nature of the end states \cite{pikulin2021protocol,aghaee2023inas}, facing, however, further doubts \cite{hess2023trivial}. Another promising path is to realize the Kitaev model in a controllable setting, such as quantum dot arrays \cite{leijnse2012parity}, cold atoms \cite{kraus2013braiding} or even available now few-qubit quantum computers \cite{mi2022noise}. This path promises to gain understanding of what is going wrong in more traditional solid state systems and how the problems might be overcome. An interesting suggestion aligned with both paths was made in Ref \cite{grusdt2016interferometric}. There it was proposed to extract topological properties of a system coupling it to a mobile impurity.

The key idea of the method is that a mobile impurity immersed in a topological background binds a few quasiparticle excitations and thus inherits topological properties of the host system \cite{grusdt2016interferometric,grusdt2019topological,pimenov2021topological,vashisht2024chiral}. These properties can be accessed in interferometric \cite{atala2013direct,grusdt2016interferometric} or transport measurements \cite{pimenov2021topological,vashisht2024chiral}. The realisation of the scheme in application to the Kitaev chain promises access to the bulk topological invariant. Polarons in a Kitaev chain also can be used to probe optically Majorana bound states \cite{mohseni2023majorana} in coupled quantum dot arrays, changing the energy scale at which the measurements have to be performed. The proposed schemes suggested are not flawless. The Ramsey interferometry suggested in Ref \cite{grusdt2019topological} cannot be easily implemented in solid state systems, such as nanowires \cite{kitaev2001unpaired} or magnetic adatoms on the surface of a superconductor \cite{nadj2014observation}. In particular, the approach \cite{grusdt2019topological} requires a specific form of the interaction between the impurity and the host system and does not work quite well with the contact interaction. The protocol suggested in Ref \cite{mohseni2023majorana} suffers from the problem of distinguishing a trivial and a topological bound state at the end of the wire. This motivated us to further investigate the polaron physics in $1d$ topological superconductors.

In this manuscript we investigate the phase diagram of a mobile impurity immersed in a $1d$ topological superconductor. In particular, we concentrate on the features which could help to distinguish between the topological and trivial phases of the host system. We find a sharp polaron-molecule phase transition in the topologically trivial background. On the other hand, the crossover between a polaron and a molecular state was observed in the topological phase. These observations are somewhat similar to ones made in \cite{qin2019polaron} regarding the trimer-polaron phase transition in a $2d$ topological superfluid. The manuscript is organized as follows. In Sec \ref{sec:model} we introduce the model and the method. In Sec \ref{sec:Results} we present the main findings of the paper. First, in Sec \ref{sec:numerical} we show the numerical phase diagram of the system and discuss its main features. In particular, we concentrate on the polaron-molecule transition and demonstrate that it has a different character depending on the phase of the host system. Thereafter, in Sec \ref{sec:exact} we consider two exactly solvable limits, illustrating the physics of the molecule-polaron transition in the Kitaev chain. Finally, in Sec \ref{sec:discussion} we summarize our findings, discuss possible experimental realisation of the system we study and suggest possible further research directions.

\section{Model and Method}
\label{sec:model}

\begin{figure}[h!]
    \centering
    \includegraphics[width=0.9\linewidth]{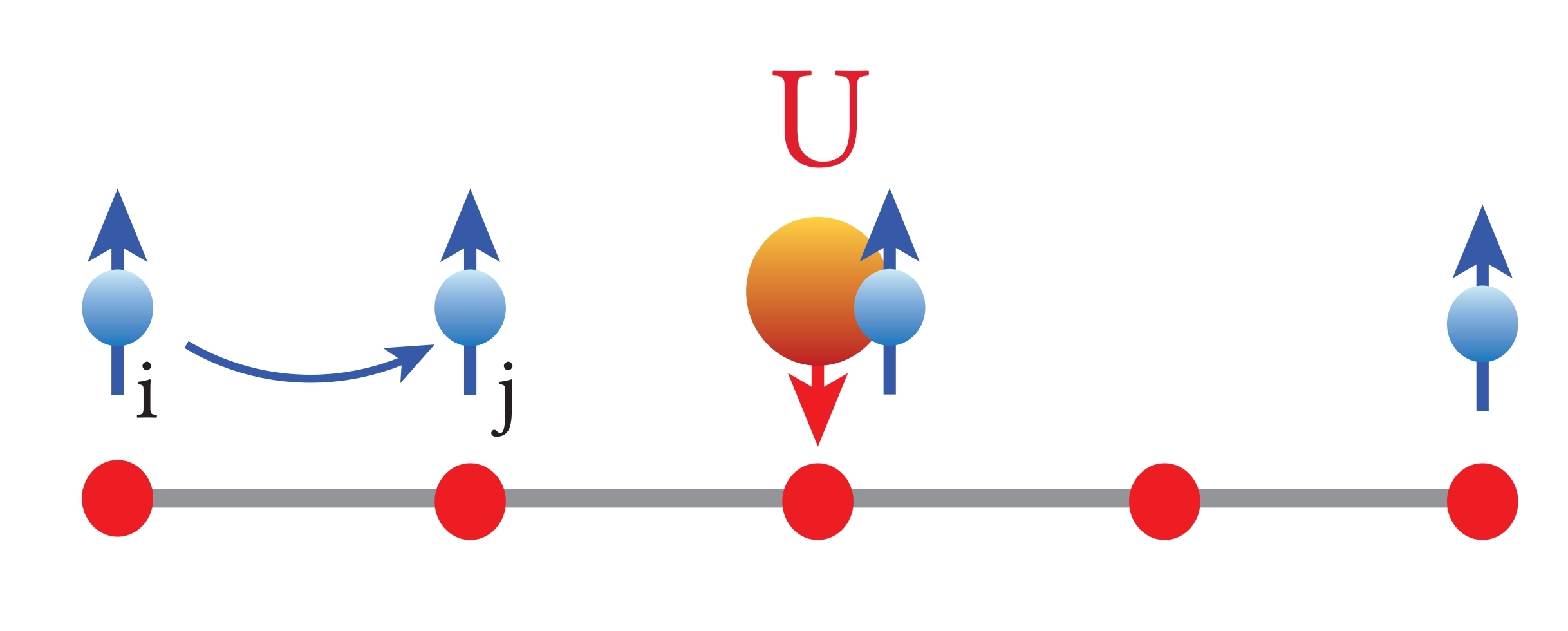}
    \caption{Schematic illustration of the system. A heavy mobile impurity (orange) is coupled to a Kitaev chain (blue) by an on-site Hubbard interaction with a strength $U$.}
    \label{fig:syst}
\end{figure}

\begin{figure*}[t]
    \includegraphics[width = 0.49\linewidth]{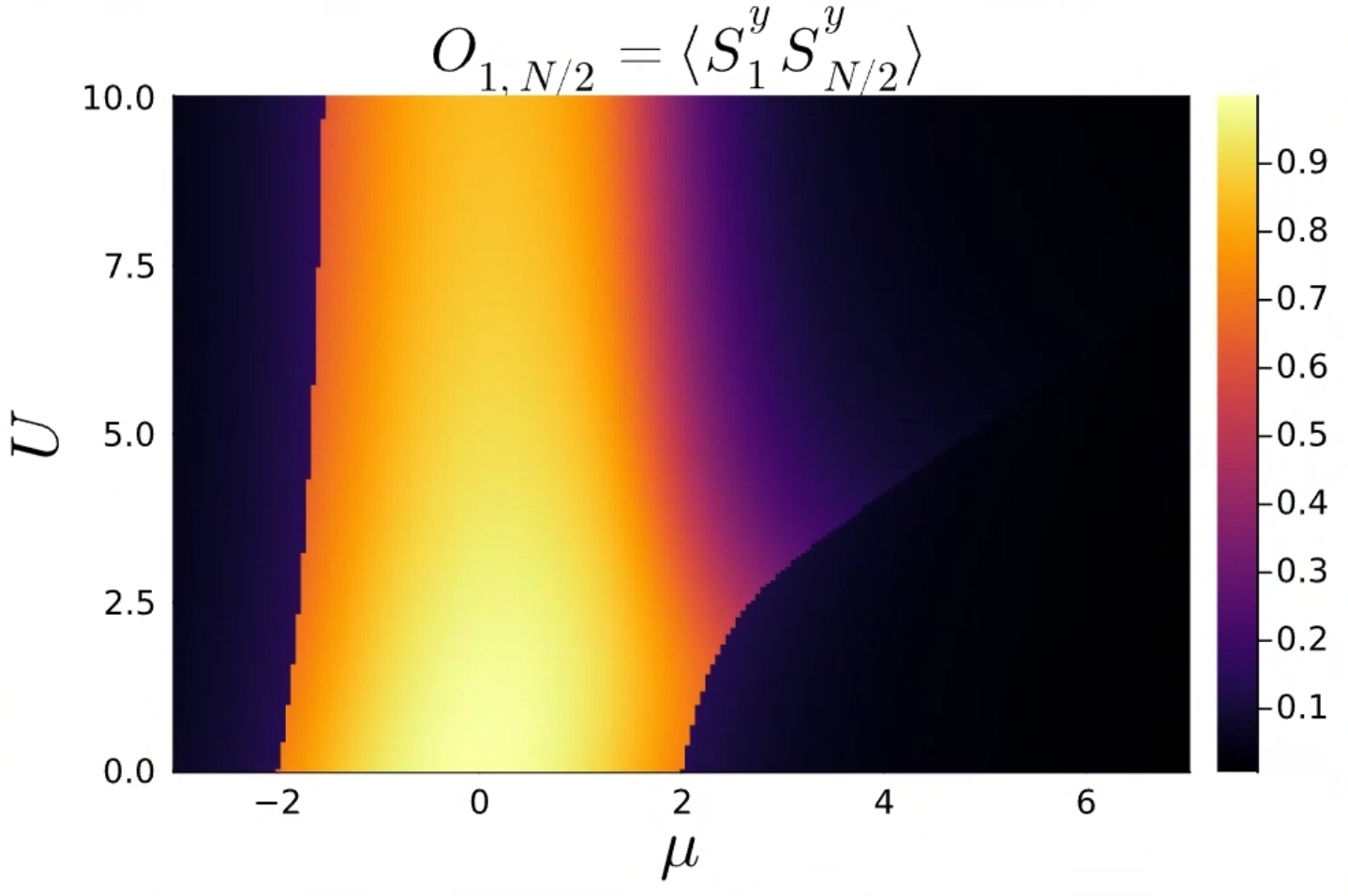}
    \includegraphics[width = 0.49\linewidth]{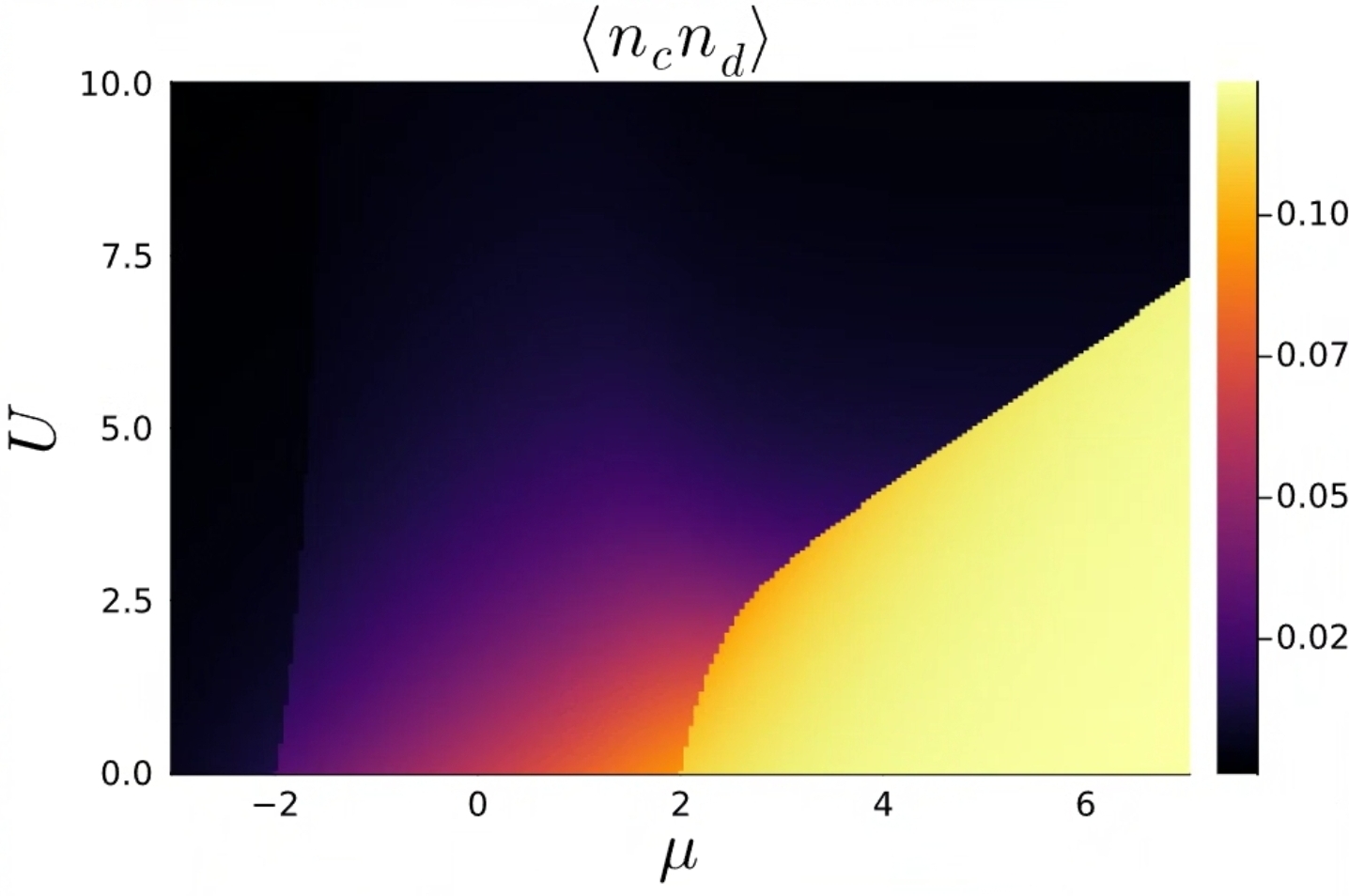}
    \caption{A phase diagrams of the closed Kitaev chain with a mobile impurity. Left figure: SOP between first and $N/2$ sites. Right figure: Molecule density correlation function; site index is dropped due to the translation invariance of the system. Parameters: $t_c=\Delta = 1, t_d = 0.1, N = 8$.}
    \label{fig:pbc_ON}
\end{figure*}
For topological superconductor we use a Kitaev chain model \cite{kitaev2001unpaired}. A mobile impurity locally interacts with the chain with a Hubbard on-site interaction (Fig~\ref{fig:syst}):
\begin{equation}\label{syst}
    H = H_{\text{Kit}} + H_\text{imp} + H_\text{int},
\end{equation}
with
\begin{equation}\label{kit}
    H_\text{Kit} = \sum_{i = 1 }^N (-t_c c^\dagger_i c_{i+1} + \Delta c^\dagger_i c_{i+1}^\dagger + h.c.) - \mu c^\dagger_i c_{i},
\end{equation}
\begin{equation}\label{imp}
    H_\text{imp} = -t_d\sum_{i=1}^N (d^\dagger_i d_{i+1} + h.c.), 
\end{equation}
\begin{equation}\label{int}
    H_\text{int} = U\sum_{i=1}^N  c^\dagger_i c_i d^\dagger_i d_i,
\end{equation}
where $t_c, t_d$ are the hopping energies of the host system and the impurity, respectively; $\Delta$ is a superconducting gap; $\mu$ is a chemical potential of the Kitaev chain; $U$ is an interaction energy; $c^\dagger_i, d^\dagger_i$ and $c_i, d_i$ are the creation and annihilation operators for the host system and the impurity correspondingly. Due to the pairing term, the number of the $c$-fermions is not fixed, while the number of impurities is set to $\sum_{j=1}^N\expval{d^\dagger_jd_j} = 1$.

The original Kitaev model Eq~\eqref{kit} features with the two phases: trivial and topological. In the topological superconducting phase, unpaired Majorana fermions must be located at the edges of the chain. One can investigate the bulk properties of the system in order to see the existence of these phases. The easiest way to do this is to close the chain and perform a discrete Fourier transform of the Hamiltonian Eq~\eqref{kit}
\begin{equation}\label{kit_fourier}
    H_\text{Kit} = \frac{1}{2}\sum_k \Psi_k^\dagger\left[-(2 t_c \cos{k} +\mu) \tau_z + 2\Delta \sin{k} \tau_y \right]\Psi_k ,
\end{equation}
where $\Psi_k^\dagger = (c^\dagger_k; c_{-k})$, $\tau_i$ are the Pauli matrices acting in the Nambu space. One can introduce a vector $\mathbf{h}_k = \left(0,2\Delta  \sin{k}, -(2t_c\cos{k} + \mu)\right)$ and calculate the winding number $\nu$ \cite{PhysRevLett.109.150408}:

\begin{equation}\label{top_triv}
    \begin{cases}
        |\mu/t_c|<2 \rightarrow \nu = 1 \text{ - topological phase} \\
        |\mu/t_c|>2 \rightarrow \nu = 0 \text{ - trivial phase}
    \end{cases}.
\end{equation}

The winding number $\nu$ is not well defined for interacting systems. Therefore we study the topology of the system using the String Order Parameter (SOP)
\cite{chitov_sop}, that can be defined as
\begin{equation}\label{sop}
    O_{ij}^y=\expval{S_i^yS^y_j},
\end{equation}
where $S_i^y$ are the spin-$\frac{1}{2}$ operators, that can be obtained from the standard Jordan-Wigner transformation:
\begin{equation}
    S_m^y = i\prod_{j=1}^{m-1}(-1)^{n_j}(c_m-c^\dagger_m).
\end{equation}

At $U\rightarrow+\infty$, the Hamiltonian Eq~\eqref{syst} can be minimized only if the interaction term gives zero contribution to the energy, so $\sum_i\expval{c^\dagger_i c_i d^\dagger_i d_i} = \sum_i\expval{n^c_in^d_i} = 0$. This means the impurity have to form a tightly bound state with the $c$-fermions. We call such a state a \textit{molecule}, not discriminating the number of $c$-fermions bound to the impurity. In the limit of weak interactions ($U\rightarrow0$), the impurity can move almost freely in the chain, but the interactions create a cloud of a few Bogoliubov quasiparticles. Thus we obtain a weakly coupled \textit{polaron} phase. One might expect that the value $\expval{n_i^c n_i^d}$ would serve as a reasonable indicator for the molecule-polaron transtion.

\section{Results}
\label{sec:Results}
\subsection{Phase diagram} 
\label{sec:numerical}
Using the Exact Diagonalization technique, we calculated the SOP (Eq \eqref{sop}) and the molecule density for the \textit{closed} Kitaev chain as functions of $U$ and $\mu$ (see Fig \ref{fig:pbc_ON}). From the SOP, we observe that as $U$ increases, the mobile impurity does not affect the boundaries of the topological phase of the Kitaev chain, and the correlations remain strong. The density-density correlator indicates no signatures of a quantum phase transition in the topological phase. Instead, there is a smooth crossover from one quantum state to another. In contrast, in the trivial phase, the density-density correlator $\expval{n^c n^d}$ experience a sharp jump at a critical value $U_c$, indicating a phase transition from the polaron state to the molecular state.

\begin{figure*}[t]
\centering
\includegraphics[width=0.49\linewidth]{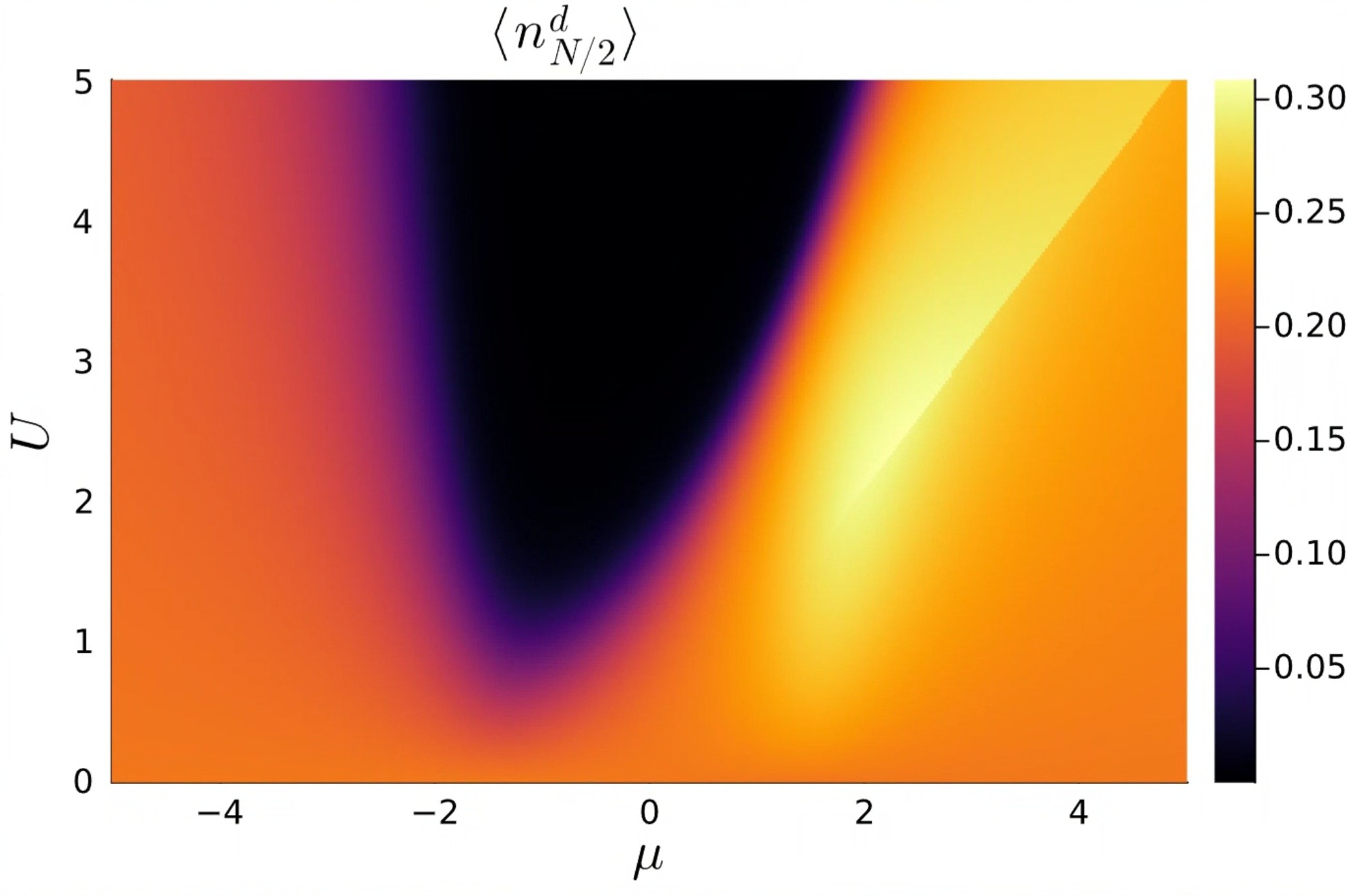}
\includegraphics[width=0.49\linewidth]{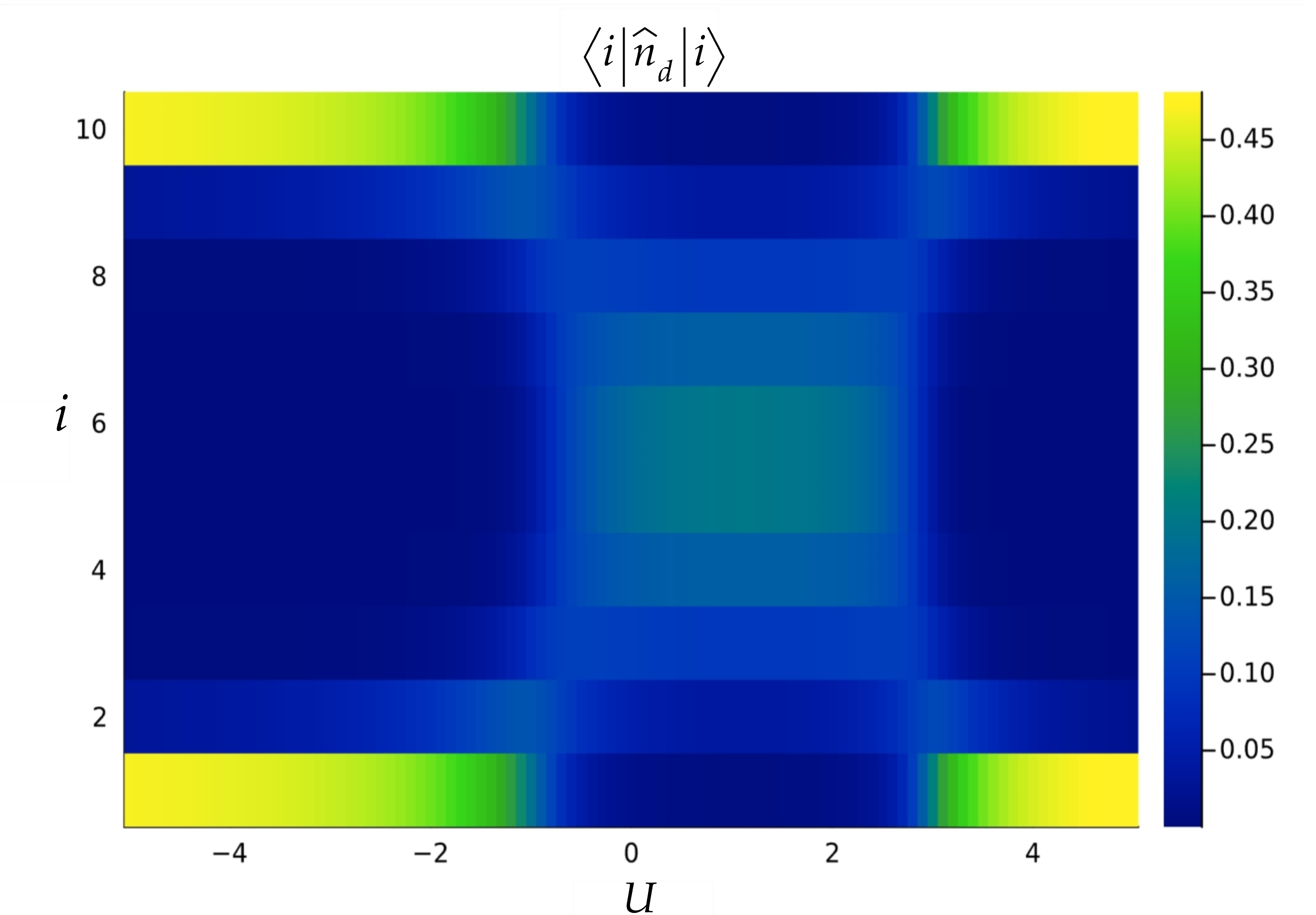}
\caption{Left: density of the impurity in the middle of the open Kitaev chain at $t_c = \Delta = 1$, $t_d = 1$, $N = 8$. Black region shows zero density (impurity edge localization). Right: impurity distribution along the chain as a function of $U$ in the topological phase ($\mu = 1$).}
\label{fig:nd_off}
\end{figure*}

For the \textit{open} chain, we also calculated the impurity density (Fig \ref{fig:nd_off}). Similar to the closed chain, there is a sharp jump in the trivial phase and a smooth crossover in the topological phase. However, an additional feature emerges: there is a region where the impurity is absent in the center of the chain, as was noticed in Ref \cite{mohseni2023majorana}. In this case, the impurity localizes at the edges (Fig \ref{fig:nd_off}: right). Importantly, the localization occurs in both the topological and trivial phases. The topological phase is characterized by the presence of two unpaired Majorana fermions localized at the edges of the chain; however, the impurity localization induces a spatial displacement of one Majorana mode to a nearest lattice site.

\subsection{Exactly Solvable Limits, Immobile Impurity}
\label{sec:exact}

In this section we explore Hamiltonian Eq~\eqref{syst} analytically to explain the differences between the polaron-molecule transitions in the topological and trivial phases. We focus on two limit cases: topological phase at $t_c = \Delta, \mu = 0$ and trivial phase at $t_c = \Delta = 0$.

\begin{figure}[h!]
\centering
\includegraphics[width=0.6\linewidth]{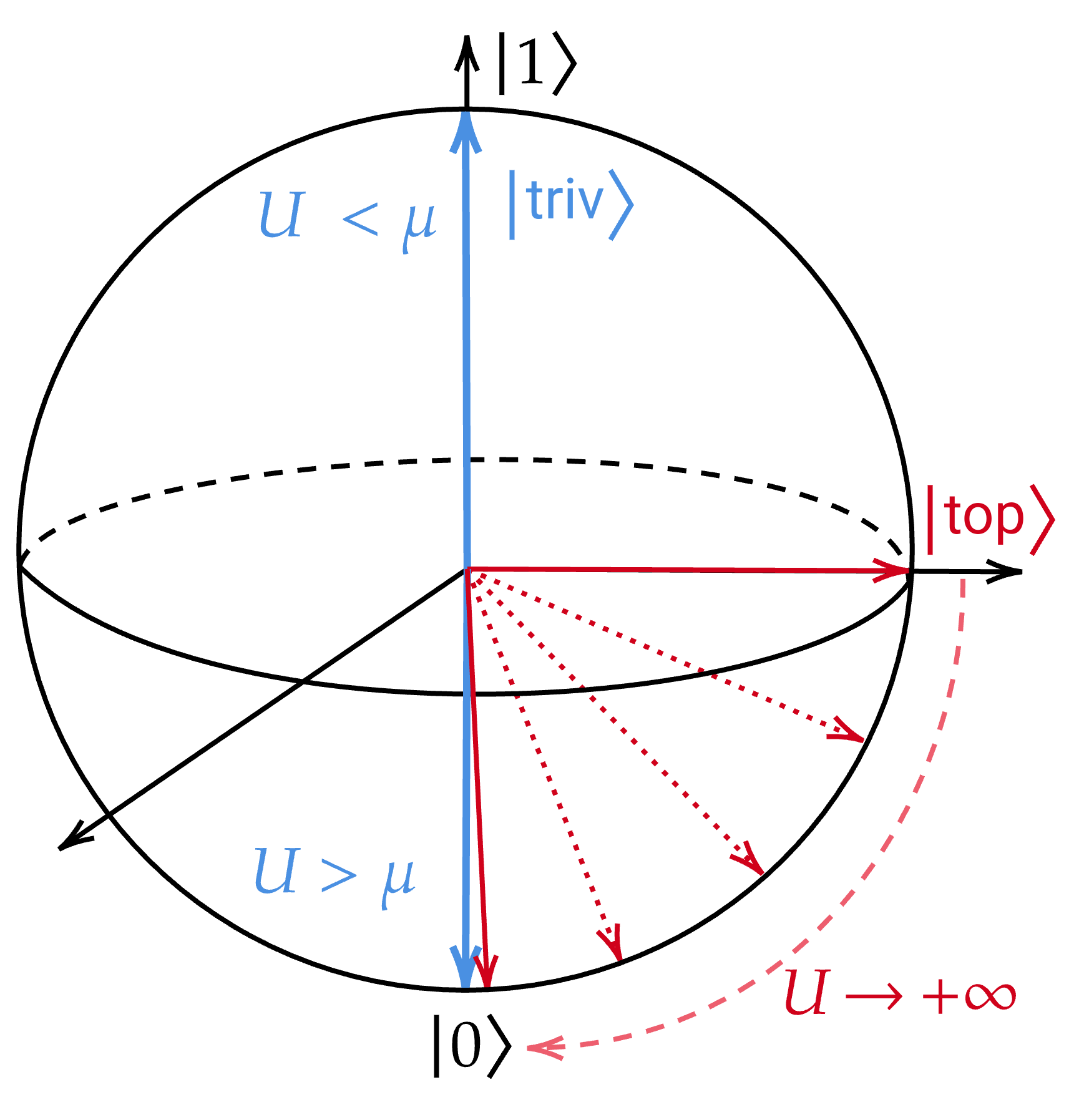}
\caption{Schematic illustration of the Bloch sphere for the site $m$ where the impurity is located. Blue lines: changes of the $c-$fermion density $n_m^c$ in the trivial phase ($t_c = \Delta = 0$) at $U = \mu$; the state jumps from $\ket{1} = \ket{\uparrow}$ to $\ket{0} = \ket{\downarrow}$. Red lines: smooth change of the density $n_m^c$ in the topological phase ($t_d = \Delta, \mu = 0$) as $U$ increases.}
\label{fig:bloch}
\end{figure}

\textit{Closed chain}: In the \textit{trivial phase} with $t_c = \Delta = 0$ and $U = 0$, Hamiltonian Eq~\eqref{syst} is simply $H = -\mu \sum_i c^\dagger_i c_i$, and one can easily find the ground state wave function $\ket{\text{triv}} = \prod_{j=1}^N c^\dagger_j \ket{\text{vac}}$. If the interaction is turned on with, the Hamiltonian will take the form
\begin{equation}
H = -\mu \sum_{i \neq m} c^\dagger_i c_i + \left(U - \mu\right) c^\dagger_m c_m,
\end{equation}
where the index $m$ is associated with the position of the impurity. Depending on the value $\lambda = U - \mu$, the ground state of the trivial phase will be:
\begin{equation} \label{triv}
\ket{\text{triv},U} = 
\begin{cases}
d^\dagger_m \prod_{j=1}^N c^\dagger_j \ket{\text{vac}}; & \text{if } \lambda < 0 \\
d^\dagger_m \prod_{j \neq m} c^\dagger_j \ket{\text{vac}}; & \text{if } \lambda > 0
\end{cases}.
\end{equation}
One can argue that $\lambda = 0$ is a point of the quantum phase transition, which must be accompanied by a sharp change of $c-$fermion density $n^c_m$ (Fig~\ref{fig:bloch}: Blue)
\begin{equation}
n_m^c = \expval{c^\dagger_m c_m } =
\begin{cases}
1; & \text{if } \lambda < 0 \\
0; & \text{if } \lambda > 0
\end{cases}.
\end{equation}

In the \textit{topological phase} with $\mu = 0$ and $t_c = \Delta$, the analysis can be simplified by introducing two Majorana representations:
\begin{equation} \label{maj}
c^\pm_j = \frac{1}{2}(\gamma_j^1 \pm i\gamma_j^2), \quad
a^\pm_j = \frac{1}{2}(\gamma_j^1 \pm i \gamma_{j-1}^2),
\end{equation}
with $\left(\gamma^1_j \right)^2 = \left(\gamma^2_j \right)^2 = 1$ and $\acomm{\gamma_j^1}{\gamma^2_i} = 0$, $\acomm{\gamma_j^1}{\gamma_i^1} = \acomm{\gamma_j^2}{\gamma_i^2} = 2\delta_{ji}$.

These representations transform the Hamiltonian in Eq~\eqref{kit} into $H_\text{Kit} = t_c \sum_j^N \left(2 a^\dagger_j a_j - 1 \right)$, with the ground state $\ket{\text{top}}$ satisfying $a^\dagger_j a_j \ket{\text{top}} = 0$ and $H_\text{Kit} \ket{\text{top}} = -N t_c \ket{\text{top}}$. When the interaction term $U$ is included and $t_d = 0$, the impurity is located on the $m$-th site ($\expval{n_m^d} = 1$) and the Hamiltonian \eqref{syst} becomes:
\begin{equation}
\begin{gathered}
H = t_c \sum_{i \neq m, m+1} \left(2 a^\dagger_i a_i - 1 \right) + 2t_c \left(a^\dagger_m a_m + a^\dagger_{m+1} a_{m+1} \right) + \\
+ \frac{U}{2} \left(a^\dagger_m a_{m+1} + a^\dagger_m a^\dagger_{m+1} + \text{h.c.} \right) - 2t_c + \frac{U}{2},
\end{gathered}
\end{equation}
Since the free and excited parts of this Hamiltonian commute with each other, we can diagonalize them separately. The excited part will be
\begin{equation}
h_\text{ex} = 
\begin{pmatrix}
0 & 0 & 0 & U/2 \\
0 & 2t_c & U/2 & 0 \\
0 & U/2 & 2t_c & 0 \\
U/2 & 0 & 0 & 4t_c
\end{pmatrix}
+ \left(\frac{U}{2} - 2t_c \right) \mathbbm{1}_{4\times4},
\end{equation}
with the minimal eigenvalue $E_\text{ex}(U) = \frac{1}{2} \left(U - \sqrt{16t_c^2 + U^2} \right)$. Since $E_{\text{ex}}$ is a smooth function of $U$, all other thermodynamic quantities must also be smooth. This observation is consistent with the results obtained from numerical calculations. The fermion density $n_m^c$ will continuously change from $0.5$ to $0$ with increasing $U$ (Fig~\ref{fig:bloch}: Red)

\textit{Open chain}: All the arguments discussed for the closed Kitaev chain can be extended to the open chain. The key question now is to determine what is the most probable position of the impurity if it is allowed to move at high $U$. Based on previous results, we expect that as $U$ increases the $\expval{n_m^d} \rightarrow 1$, where $m$ is the position of the impurity. If $m = 1$ or $m = N$ and $U \rightarrow +\infty$, the minimal energy of Eq~\eqref{syst} in the Majorana representation \eqref{maj}:
\begin{equation} \label{Uinf_spin}
E_{1;N} = -\sum_{i \neq 1} \left(4t_c \expval{i\gamma^1_{i+1} \gamma^2_i} + \mu \expval{i\gamma^1_i \gamma^2_i} \right) = E_\text{Kit}(N-1),
\end{equation}
where $E_\text{Kit}(N-1)$ is the energy of the Kitaev chain with $N-1$ sites. If $m \neq 1, N$, the minimal energy will be $E_{m \neq 1,N} = E_\text{Kit}(m-1) + E_\text{Kit}(N-m)$. If we compare these two energies, one can find that
\begin{equation}
E_\text{Kit}(N-1) < E_\text{Kit}(m-1) + E_\text{Kit}(N-m).
\end{equation}
This means, at high $U$ the impurity will be localized at the edges of the Kitaev chain in both topological and trivial phases.


\section{Discussion}
\label{sec:discussion}

We considered phases of a heavy mobile impurity coupled to a $1d$ topological superconductor. We found that a polaron-molecule transition of the impurity can be used as a signature of the phase of the host system. One might expect that the physics we discuss in the manuscript would survive in the presence of disorder. Impurity binding of a $c$-fermionic hole in a trivial insulator is accompanied by a change of parity, leading to the phase transition. On the other hand, a strongly coupled state in the topological phase requires annihilation of two fermions as the interactions grow. Thus, the system undergoes a crossover. However, further investigation of the effects of disorder on the mobile impurity is a natural subject for future studies. Let us finally discuss possible routes to experimental realisation of the considered system. Most directly, our results are applicable to artificial systems: cold atoms as suggested in Ref \cite{grusdt2019topological} and quantum dot arrays \cite{mohseni2023majorana}. However, recent progress in the field \cite{sidler2017fermi} gives hope for using the method we propose in semiconductor systems.    

\section*{Acknowledgments}
The work was supported by Rosatom in the framework of the Roadmap for Quantum Computing (Contract No. 868-1.3-15/15-2021 dated October 5). This research was supported in part through computational resources of the HPC facilities at HSE University. M. Bachovadinov thanks the Basic Research Program of HSE for the provided support.

\appendix{}

\section{Correlations in the case of even and odd $N$}

\begin{figure}[h!]
    \centering
    \includegraphics[width=1\linewidth]{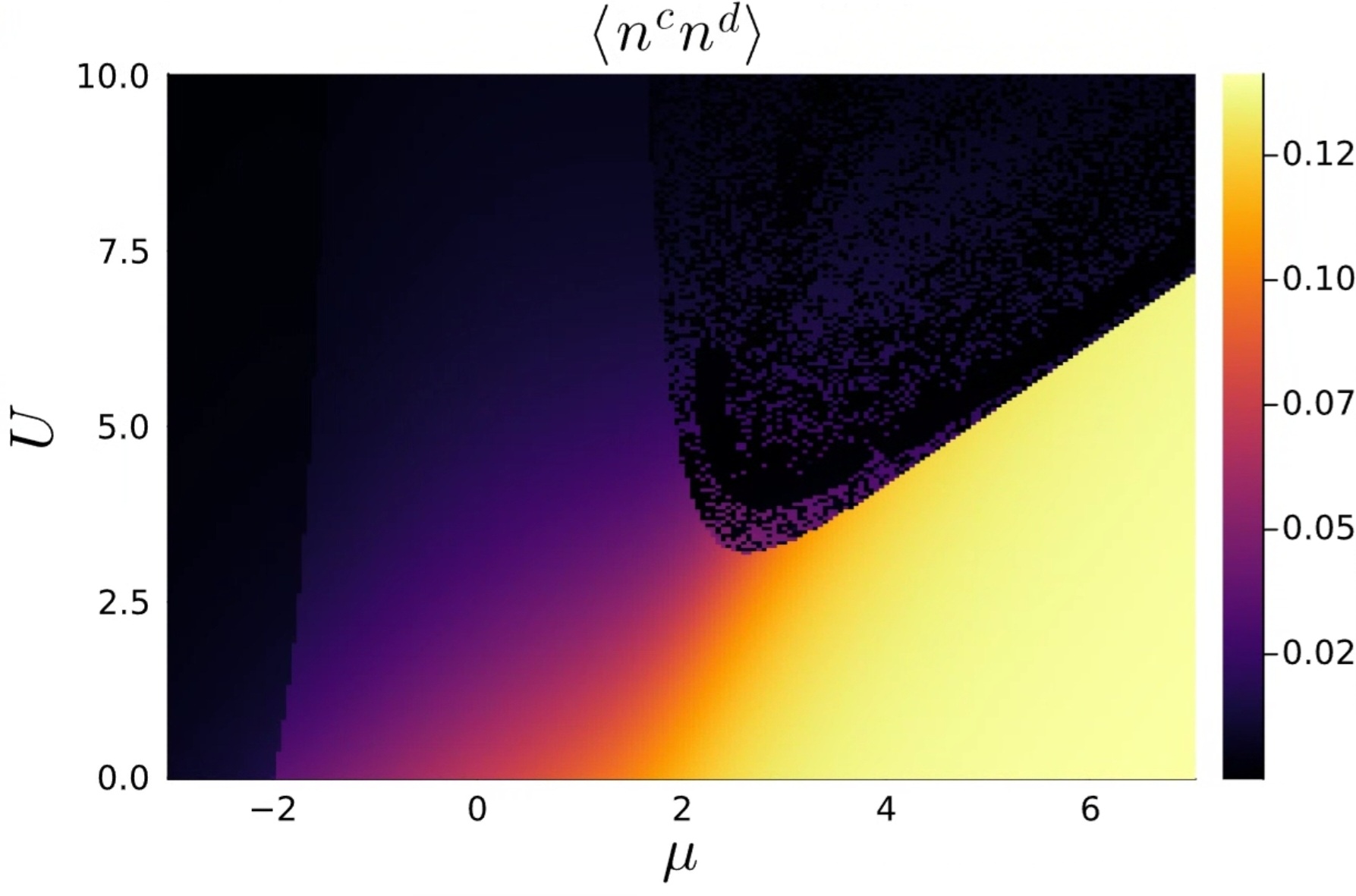}
    \caption{Molecule density correlation function for odd $N$. Parameters: $t_c = \Delta = 1, t_d = 0.1, N = 9$}
    \label{fig:corr_odd}
\end{figure}

The numerical calculations in section \ref{sec:numerical} were performed for even $N$. Here we want to clarify the differences between the even and odd number of sites. For odd $N$, the $\expval{n^c n^d}$ correlation function on the closed chain (Fig \ref{fig:corr_odd}) behaves differently compared with the even case.   

To understand why this happens, we take a look at the original Kitaev model ($U=0$). One can easily find the analytical expression for the on-site electron density:
\begin{equation}\label{a1}
    \expval{n^c} = \frac{1}{2N}\sum_{k} \left(1 -\frac{\varepsilon_k}{E_k}\right),
\end{equation}
where $\Delta k = \frac{2\pi}{N}$, $E_k = \sqrt{(2t_c\cos{k}+\mu)^2 + 4\Delta^2 \sin^2{k}}$ is the spectrum of the Kitaev chain, and $\varepsilon_k = -2t_c\cos{k} -\mu$. We will focus on the properties of the function $g_k(\mu) = \frac{1}{2N}\frac{\varepsilon_k}{E_k}$. At $k=\pi$ it takes the form:
\begin{equation}\label{gap}
    g_\pi(\mu) = \frac{1}{2N}\text{sign}(\mu-2)= \begin{cases}
        \frac{1}{2N}, \text{ if } \mu>2 \\
         -\frac{1}{2N}, \text{ if } \mu<2 
    \end{cases},
\end{equation}
so $g_k(\mu)$ has a discontinuity that depends on $N$ at $k=\pi$. Such behaviour also appears at $k=0$, $\mu=-2$, which can be written as:
\begin{equation}\label{a3}
    g_0(-2) = - g_\pi(2).
\end{equation}

But Eq\eqref{a3} can't be reached if $N$ is odd, because the momentum quantization is $k_j = \frac{2\pi}{N}j$ with $j=0,...,N-1$, and $k$ can't be equal to $\pi$ at low $N$. In this case, there is no discontinuity in $g_k(\mu)$ at $\mu=2$, as we see in Fig \ref{fig:corr_odd} at $U=0$.
For $U\neq 0$, the momentum quantization doesn't change, so one should expect analogous behaviour of the correlators at low $N$. 

\begin{figure}[h!]
    \centering
    \includegraphics[width=0.99\linewidth]{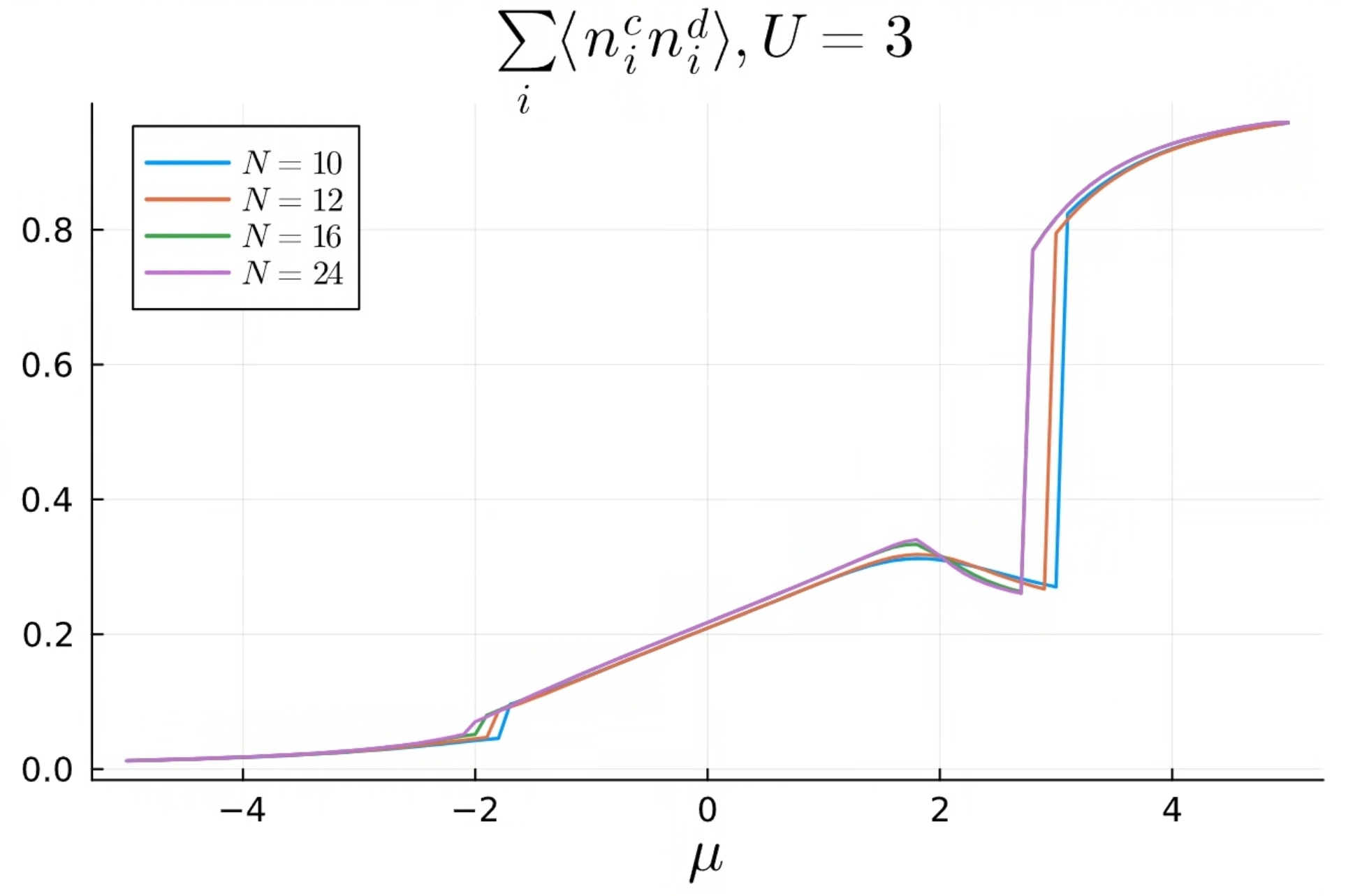}
    \caption{Molecule density $\sum_i^N\expval{n_i^c n_i^d}$ at fixed $U=3$ for the different number of sites $N$.}
    \label{fig:corr_u=3}
\end{figure}

While the discontinuity described by Eq~\eqref{gap} vanishes for large system sizes $N$, there is still a question about the jump at the quantum phase transition point. 
Fig~\ref{fig:corr_u=3} shows the molecular density as a function of 
the chemical potential $\mu$. Notably, for system sizes $N \geq 16$, the discontinuity associated with the phase transition remains constant, suggesting that this characteristic behaviour persists in the thermodynamic limit.

\newpage


\begin{thebibliography}{33}%
\makeatletter
\providecommand \@ifxundefined [1]{%
 \@ifx{#1\undefined}
}%
\providecommand \@ifnum [1]{%
 \ifnum #1\expandafter \@firstoftwo
 \else \expandafter \@secondoftwo
 \fi
}%
\providecommand \@ifx [1]{%
 \ifx #1\expandafter \@firstoftwo
 \else \expandafter \@secondoftwo
 \fi
}%
\providecommand \natexlab [1]{#1}%
\providecommand \enquote  [1]{``#1''}%
\providecommand \bibnamefont  [1]{#1}%
\providecommand \bibfnamefont [1]{#1}%
\providecommand \citenamefont [1]{#1}%
\providecommand \href@noop [0]{\@secondoftwo}%
\providecommand \href [0]{\begingroup \@sanitize@url \@href}%
\providecommand \@href[1]{\@@startlink{#1}\@@href}%
\providecommand \@@href[1]{\endgroup#1\@@endlink}%
\providecommand \@sanitize@url [0]{\catcode `\\12\catcode `\$12\catcode `\&12\catcode `\#12\catcode `\^12\catcode `\_12\catcode `\%12\relax}%
\providecommand \@@startlink[1]{}%
\providecommand \@@endlink[0]{}%
\providecommand \url  [0]{\begingroup\@sanitize@url \@url }%
\providecommand \@url [1]{\endgroup\@href {#1}{\urlprefix }}%
\providecommand \urlprefix  [0]{URL }%
\providecommand \Eprint [0]{\href }%
\providecommand \doibase [0]{https://doi.org/}%
\providecommand \selectlanguage [0]{\@gobble}%
\providecommand \bibinfo  [0]{\@secondoftwo}%
\providecommand \bibfield  [0]{\@secondoftwo}%
\providecommand \translation [1]{[#1]}%
\providecommand \BibitemOpen [0]{}%
\providecommand \bibitemStop [0]{}%
\providecommand \bibitemNoStop [0]{.\EOS\space}%
\providecommand \EOS [0]{\spacefactor3000\relax}%
\providecommand \BibitemShut  [1]{\csname bibitem#1\endcsname}%
\let\auto@bib@innerbib\@empty
\bibitem [{\citenamefont {Klitzing}\ \emph {et~al.}(1980)\citenamefont {Klitzing}, \citenamefont {Dorda},\ and\ \citenamefont {Pepper}}]{klitzing1980new}%
  \BibitemOpen
  \bibfield  {author} {\bibinfo {author} {\bibfnamefont {K.~v.}\ \bibnamefont {Klitzing}}, \bibinfo {author} {\bibfnamefont {G.}~\bibnamefont {Dorda}},\ and\ \bibinfo {author} {\bibfnamefont {M.}~\bibnamefont {Pepper}},\ }\bibfield  {title} {\bibinfo {title} {New method for high-accuracy determination of the fine-structure constant based on quantized hall resistance},\ }\href@noop {} {\bibfield  {journal} {\bibinfo  {journal} {Physical review letters}\ }\textbf {\bibinfo {volume} {45}},\ \bibinfo {pages} {494} (\bibinfo {year} {1980})}\BibitemShut {NoStop}%
\bibitem [{\citenamefont {Halperin}(1982)}]{halperin1982quantized}%
  \BibitemOpen
  \bibfield  {author} {\bibinfo {author} {\bibfnamefont {B.~I.}\ \bibnamefont {Halperin}},\ }\bibfield  {title} {\bibinfo {title} {Quantized hall conductance, current-carrying edge states, and the existence of extended states in a two-dimensional disordered potential},\ }\href@noop {} {\bibfield  {journal} {\bibinfo  {journal} {Physical review B}\ }\textbf {\bibinfo {volume} {25}},\ \bibinfo {pages} {2185} (\bibinfo {year} {1982})}\BibitemShut {NoStop}%
\bibitem [{\citenamefont {Arovas}\ \emph {et~al.}(1984)\citenamefont {Arovas}, \citenamefont {Schrieffer},\ and\ \citenamefont {Wilczek}}]{arovas1984fractional}%
  \BibitemOpen
  \bibfield  {author} {\bibinfo {author} {\bibfnamefont {D.}~\bibnamefont {Arovas}}, \bibinfo {author} {\bibfnamefont {J.~R.}\ \bibnamefont {Schrieffer}},\ and\ \bibinfo {author} {\bibfnamefont {F.}~\bibnamefont {Wilczek}},\ }\bibfield  {title} {\bibinfo {title} {Fractional statistics and the quantum hall effect},\ }\href@noop {} {\bibfield  {journal} {\bibinfo  {journal} {Physical review letters}\ }\textbf {\bibinfo {volume} {53}},\ \bibinfo {pages} {722} (\bibinfo {year} {1984})}\BibitemShut {NoStop}%
\bibitem [{\citenamefont {Fu}(2011)}]{fu2011topological}%
  \BibitemOpen
  \bibfield  {author} {\bibinfo {author} {\bibfnamefont {L.}~\bibnamefont {Fu}},\ }\bibfield  {title} {\bibinfo {title} {Topological crystalline insulators},\ }\href@noop {} {\bibfield  {journal} {\bibinfo  {journal} {Physical review letters}\ }\textbf {\bibinfo {volume} {106}},\ \bibinfo {pages} {106802} (\bibinfo {year} {2011})}\BibitemShut {NoStop}%
\bibitem [{\citenamefont {Yu}\ \emph {et~al.}(2021)\citenamefont {Yu}, \citenamefont {Chen}, \citenamefont {Gomanko}, \citenamefont {Badawy}, \citenamefont {Bakkers}, \citenamefont {Zuo}, \citenamefont {Mourik},\ and\ \citenamefont {Frolov}}]{yu2021non}%
  \BibitemOpen
  \bibfield  {author} {\bibinfo {author} {\bibfnamefont {P.}~\bibnamefont {Yu}}, \bibinfo {author} {\bibfnamefont {J.}~\bibnamefont {Chen}}, \bibinfo {author} {\bibfnamefont {M.}~\bibnamefont {Gomanko}}, \bibinfo {author} {\bibfnamefont {G.}~\bibnamefont {Badawy}}, \bibinfo {author} {\bibfnamefont {E.}~\bibnamefont {Bakkers}}, \bibinfo {author} {\bibfnamefont {K.}~\bibnamefont {Zuo}}, \bibinfo {author} {\bibfnamefont {V.}~\bibnamefont {Mourik}},\ and\ \bibinfo {author} {\bibfnamefont {S.}~\bibnamefont {Frolov}},\ }\bibfield  {title} {\bibinfo {title} {Non-majorana states yield nearly quantized conductance in proximatized nanowires},\ }\href@noop {} {\bibfield  {journal} {\bibinfo  {journal} {Nature Physics}\ }\textbf {\bibinfo {volume} {17}},\ \bibinfo {pages} {482} (\bibinfo {year} {2021})}\BibitemShut {NoStop}%
\bibitem [{\citenamefont {Kitaev}(2001)}]{kitaev2001unpaired}%
  \BibitemOpen
  \bibfield  {author} {\bibinfo {author} {\bibfnamefont {A.~Y.}\ \bibnamefont {Kitaev}},\ }\bibfield  {title} {\bibinfo {title} {Unpaired majorana fermions in quantum wires},\ }\href@noop {} {\bibfield  {journal} {\bibinfo  {journal} {Physics-uspekhi}\ }\textbf {\bibinfo {volume} {44}},\ \bibinfo {pages} {131} (\bibinfo {year} {2001})}\BibitemShut {NoStop}%
\bibitem [{\citenamefont {Frolov}(2021)}]{frolov2021quantum}%
  \BibitemOpen
  \bibfield  {author} {\bibinfo {author} {\bibfnamefont {S.}~\bibnamefont {Frolov}},\ }\bibfield  {title} {\bibinfo {title} {Quantum computing’s reproducibility crisis: Majorana fermions},\ }\href@noop {} {\bibfield  {journal} {\bibinfo  {journal} {Nature}\ }\textbf {\bibinfo {volume} {592}},\ \bibinfo {pages} {350} (\bibinfo {year} {2021})}\BibitemShut {NoStop}%
\bibitem [{\citenamefont {Akhmerov}\ \emph {et~al.}(2011)\citenamefont {Akhmerov}, \citenamefont {Dahlhaus}, \citenamefont {Hassler}, \citenamefont {Wimmer},\ and\ \citenamefont {Beenakker}}]{akhmerov2011quantized}%
  \BibitemOpen
  \bibfield  {author} {\bibinfo {author} {\bibfnamefont {A.}~\bibnamefont {Akhmerov}}, \bibinfo {author} {\bibfnamefont {J.}~\bibnamefont {Dahlhaus}}, \bibinfo {author} {\bibfnamefont {F.}~\bibnamefont {Hassler}}, \bibinfo {author} {\bibfnamefont {M.}~\bibnamefont {Wimmer}},\ and\ \bibinfo {author} {\bibfnamefont {C.}~\bibnamefont {Beenakker}},\ }\bibfield  {title} {\bibinfo {title} {Quantized conductance at the majorana phase transition in a disordered<? format?> superconducting wire},\ }\href@noop {} {\bibfield  {journal} {\bibinfo  {journal} {Physical review letters}\ }\textbf {\bibinfo {volume} {106}},\ \bibinfo {pages} {057001} (\bibinfo {year} {2011})}\BibitemShut {NoStop}%
\bibitem [{\citenamefont {Mourik}\ \emph {et~al.}(2012)\citenamefont {Mourik}, \citenamefont {Zuo}, \citenamefont {Frolov}, \citenamefont {Plissard}, \citenamefont {Bakkers},\ and\ \citenamefont {Kouwenhoven}}]{mourik2012signatures}%
  \BibitemOpen
  \bibfield  {author} {\bibinfo {author} {\bibfnamefont {V.}~\bibnamefont {Mourik}}, \bibinfo {author} {\bibfnamefont {K.}~\bibnamefont {Zuo}}, \bibinfo {author} {\bibfnamefont {S.~M.}\ \bibnamefont {Frolov}}, \bibinfo {author} {\bibfnamefont {S.}~\bibnamefont {Plissard}}, \bibinfo {author} {\bibfnamefont {E.~P.}\ \bibnamefont {Bakkers}},\ and\ \bibinfo {author} {\bibfnamefont {L.~P.}\ \bibnamefont {Kouwenhoven}},\ }\bibfield  {title} {\bibinfo {title} {Signatures of majorana fermions in hybrid superconductor-semiconductor nanowire devices},\ }\href@noop {} {\bibfield  {journal} {\bibinfo  {journal} {Science}\ }\textbf {\bibinfo {volume} {336}},\ \bibinfo {pages} {1003} (\bibinfo {year} {2012})}\BibitemShut {NoStop}%
\bibitem [{\citenamefont {G{\"u}l}\ \emph {et~al.}(2018)\citenamefont {G{\"u}l}, \citenamefont {Zhang}, \citenamefont {Bommer}, \citenamefont {de~Moor}, \citenamefont {Car}, \citenamefont {Plissard}, \citenamefont {Bakkers}, \citenamefont {Geresdi}, \citenamefont {Watanabe}, \citenamefont {Taniguchi} \emph {et~al.}}]{gul2018ballistic}%
  \BibitemOpen
  \bibfield  {author} {\bibinfo {author} {\bibfnamefont {{\"O}.}~\bibnamefont {G{\"u}l}}, \bibinfo {author} {\bibfnamefont {H.}~\bibnamefont {Zhang}}, \bibinfo {author} {\bibfnamefont {J.~D.}\ \bibnamefont {Bommer}}, \bibinfo {author} {\bibfnamefont {M.~W.}\ \bibnamefont {de~Moor}}, \bibinfo {author} {\bibfnamefont {D.}~\bibnamefont {Car}}, \bibinfo {author} {\bibfnamefont {S.~R.}\ \bibnamefont {Plissard}}, \bibinfo {author} {\bibfnamefont {E.~P.}\ \bibnamefont {Bakkers}}, \bibinfo {author} {\bibfnamefont {A.}~\bibnamefont {Geresdi}}, \bibinfo {author} {\bibfnamefont {K.}~\bibnamefont {Watanabe}}, \bibinfo {author} {\bibfnamefont {T.}~\bibnamefont {Taniguchi}}, \emph {et~al.},\ }\bibfield  {title} {\bibinfo {title} {Ballistic majorana nanowire devices},\ }\href@noop {} {\bibfield  {journal} {\bibinfo  {journal} {Nature nanotechnology}\ }\textbf {\bibinfo {volume} {13}},\ \bibinfo {pages} {192} (\bibinfo {year} {2018})}\BibitemShut {NoStop}%
\bibitem [{\citenamefont {Zhang}\ \emph {et~al.}(2018)\citenamefont {Zhang}, \citenamefont {Liu}, \citenamefont {Gazibegovic}, \citenamefont {Xu}, \citenamefont {Logan}, \citenamefont {Wang}, \citenamefont {Van~Loo}, \citenamefont {Bommer}, \citenamefont {De~Moor}, \citenamefont {Car} \emph {et~al.}}]{zhang2018retracted}%
  \BibitemOpen
  \bibfield  {author} {\bibinfo {author} {\bibfnamefont {H.}~\bibnamefont {Zhang}}, \bibinfo {author} {\bibfnamefont {C.-X.}\ \bibnamefont {Liu}}, \bibinfo {author} {\bibfnamefont {S.}~\bibnamefont {Gazibegovic}}, \bibinfo {author} {\bibfnamefont {D.}~\bibnamefont {Xu}}, \bibinfo {author} {\bibfnamefont {J.~A.}\ \bibnamefont {Logan}}, \bibinfo {author} {\bibfnamefont {G.}~\bibnamefont {Wang}}, \bibinfo {author} {\bibfnamefont {N.}~\bibnamefont {Van~Loo}}, \bibinfo {author} {\bibfnamefont {J.~D.}\ \bibnamefont {Bommer}}, \bibinfo {author} {\bibfnamefont {M.~W.}\ \bibnamefont {De~Moor}}, \bibinfo {author} {\bibfnamefont {D.}~\bibnamefont {Car}}, \emph {et~al.},\ }\bibfield  {title} {\bibinfo {title} {Retracted article: Quantized majorana conductance},\ }\href@noop {} {\bibfield  {journal} {\bibinfo  {journal} {Nature}\ }\textbf {\bibinfo {volume} {556}},\ \bibinfo {pages} {74} (\bibinfo {year} {2018})}\BibitemShut {NoStop}%
\bibitem [{\citenamefont {Nadj-Perge}\ \emph {et~al.}(2014)\citenamefont {Nadj-Perge}, \citenamefont {Drozdov}, \citenamefont {Li}, \citenamefont {Chen}, \citenamefont {Jeon}, \citenamefont {Seo}, \citenamefont {MacDonald}, \citenamefont {Bernevig},\ and\ \citenamefont {Yazdani}}]{nadj2014observation}%
  \BibitemOpen
  \bibfield  {author} {\bibinfo {author} {\bibfnamefont {S.}~\bibnamefont {Nadj-Perge}}, \bibinfo {author} {\bibfnamefont {I.~K.}\ \bibnamefont {Drozdov}}, \bibinfo {author} {\bibfnamefont {J.}~\bibnamefont {Li}}, \bibinfo {author} {\bibfnamefont {H.}~\bibnamefont {Chen}}, \bibinfo {author} {\bibfnamefont {S.}~\bibnamefont {Jeon}}, \bibinfo {author} {\bibfnamefont {J.}~\bibnamefont {Seo}}, \bibinfo {author} {\bibfnamefont {A.~H.}\ \bibnamefont {MacDonald}}, \bibinfo {author} {\bibfnamefont {B.~A.}\ \bibnamefont {Bernevig}},\ and\ \bibinfo {author} {\bibfnamefont {A.}~\bibnamefont {Yazdani}},\ }\bibfield  {title} {\bibinfo {title} {Observation of majorana fermions in ferromagnetic atomic chains on a superconductor},\ }\href@noop {} {\bibfield  {journal} {\bibinfo  {journal} {Science}\ }\textbf {\bibinfo {volume} {346}},\ \bibinfo {pages} {602} (\bibinfo {year} {2014})}\BibitemShut {NoStop}%
\bibitem [{\citenamefont {Lee}\ \emph {et~al.}(2012)\citenamefont {Lee}, \citenamefont {Jiang}, \citenamefont {Aguado}, \citenamefont {Katsaros}, \citenamefont {Lieber},\ and\ \citenamefont {De~Franceschi}}]{lee2012zero}%
  \BibitemOpen
  \bibfield  {author} {\bibinfo {author} {\bibfnamefont {E.~J.}\ \bibnamefont {Lee}}, \bibinfo {author} {\bibfnamefont {X.}~\bibnamefont {Jiang}}, \bibinfo {author} {\bibfnamefont {R.}~\bibnamefont {Aguado}}, \bibinfo {author} {\bibfnamefont {G.}~\bibnamefont {Katsaros}}, \bibinfo {author} {\bibfnamefont {C.~M.}\ \bibnamefont {Lieber}},\ and\ \bibinfo {author} {\bibfnamefont {S.}~\bibnamefont {De~Franceschi}},\ }\bibfield  {title} {\bibinfo {title} {Zero-bias anomaly in a nanowire quantum dot coupled to superconductors},\ }\href@noop {} {\bibfield  {journal} {\bibinfo  {journal} {Physical review letters}\ }\textbf {\bibinfo {volume} {109}},\ \bibinfo {pages} {186802} (\bibinfo {year} {2012})}\BibitemShut {NoStop}%
\bibitem [{\citenamefont {Sau}\ and\ \citenamefont {Brydon}(2015)}]{sau2015bound}%
  \BibitemOpen
  \bibfield  {author} {\bibinfo {author} {\bibfnamefont {J.~D.}\ \bibnamefont {Sau}}\ and\ \bibinfo {author} {\bibfnamefont {P.}~\bibnamefont {Brydon}},\ }\bibfield  {title} {\bibinfo {title} {Bound states of a ferromagnetic wire in a superconductor},\ }\href@noop {} {\bibfield  {journal} {\bibinfo  {journal} {Physical review letters}\ }\textbf {\bibinfo {volume} {115}},\ \bibinfo {pages} {127003} (\bibinfo {year} {2015})}\BibitemShut {NoStop}%
\bibitem [{\citenamefont {Frolov}\ and\ \citenamefont {Mourik}(2024)}]{frolov2024comment}%
  \BibitemOpen
  \bibfield  {author} {\bibinfo {author} {\bibfnamefont {S.}~\bibnamefont {Frolov}}\ and\ \bibinfo {author} {\bibfnamefont {V.}~\bibnamefont {Mourik}},\ }\bibfield  {title} {\bibinfo {title} {Comment on" ballistic majorana nanowire devices" by gul et al. nature nanotechnology 2018},\ }\href@noop {} {\bibfield  {journal} {\bibinfo  {journal} {arXiv preprint arXiv:2407.18623}\ } (\bibinfo {year} {2024})}\BibitemShut {NoStop}%
\bibitem [{\citenamefont {Hess}\ \emph {et~al.}(2022)\citenamefont {Hess}, \citenamefont {Legg}, \citenamefont {Loss},\ and\ \citenamefont {Klinovaja}}]{hess2022prevalence}%
  \BibitemOpen
  \bibfield  {author} {\bibinfo {author} {\bibfnamefont {R.}~\bibnamefont {Hess}}, \bibinfo {author} {\bibfnamefont {H.~F.}\ \bibnamefont {Legg}}, \bibinfo {author} {\bibfnamefont {D.}~\bibnamefont {Loss}},\ and\ \bibinfo {author} {\bibfnamefont {J.}~\bibnamefont {Klinovaja}},\ }\bibfield  {title} {\bibinfo {title} {Prevalence of trivial zero-energy subgap states in nonuniform helical spin chains on the surface of superconductors},\ }\href@noop {} {\bibfield  {journal} {\bibinfo  {journal} {Physical Review B}\ }\textbf {\bibinfo {volume} {106}},\ \bibinfo {pages} {104503} (\bibinfo {year} {2022})}\BibitemShut {NoStop}%
\bibitem [{\citenamefont {Yazdani}\ \emph {et~al.}(2023)\citenamefont {Yazdani}, \citenamefont {Von~Oppen}, \citenamefont {Halperin},\ and\ \citenamefont {Yacoby}}]{yazdani2023hunting}%
  \BibitemOpen
  \bibfield  {author} {\bibinfo {author} {\bibfnamefont {A.}~\bibnamefont {Yazdani}}, \bibinfo {author} {\bibfnamefont {F.}~\bibnamefont {Von~Oppen}}, \bibinfo {author} {\bibfnamefont {B.~I.}\ \bibnamefont {Halperin}},\ and\ \bibinfo {author} {\bibfnamefont {A.}~\bibnamefont {Yacoby}},\ }\bibfield  {title} {\bibinfo {title} {Hunting for majoranas},\ }\href@noop {} {\bibfield  {journal} {\bibinfo  {journal} {Science}\ }\textbf {\bibinfo {volume} {380}},\ \bibinfo {pages} {eade0850} (\bibinfo {year} {2023})}\BibitemShut {NoStop}%
\bibitem [{\citenamefont {Pikulin}\ \emph {et~al.}(2021)\citenamefont {Pikulin}, \citenamefont {van Heck}, \citenamefont {Karzig}, \citenamefont {Martinez}, \citenamefont {Nijholt}, \citenamefont {Laeven}, \citenamefont {Winkler}, \citenamefont {Watson}, \citenamefont {Heedt}, \citenamefont {Temurhan} \emph {et~al.}}]{pikulin2021protocol}%
  \BibitemOpen
  \bibfield  {author} {\bibinfo {author} {\bibfnamefont {D.~I.}\ \bibnamefont {Pikulin}}, \bibinfo {author} {\bibfnamefont {B.}~\bibnamefont {van Heck}}, \bibinfo {author} {\bibfnamefont {T.}~\bibnamefont {Karzig}}, \bibinfo {author} {\bibfnamefont {E.~A.}\ \bibnamefont {Martinez}}, \bibinfo {author} {\bibfnamefont {B.}~\bibnamefont {Nijholt}}, \bibinfo {author} {\bibfnamefont {T.}~\bibnamefont {Laeven}}, \bibinfo {author} {\bibfnamefont {G.~W.}\ \bibnamefont {Winkler}}, \bibinfo {author} {\bibfnamefont {J.~D.}\ \bibnamefont {Watson}}, \bibinfo {author} {\bibfnamefont {S.}~\bibnamefont {Heedt}}, \bibinfo {author} {\bibfnamefont {M.}~\bibnamefont {Temurhan}}, \emph {et~al.},\ }\bibfield  {title} {\bibinfo {title} {Protocol to identify a topological superconducting phase in a three-terminal device},\ }\href@noop {} {\bibfield  {journal} {\bibinfo  {journal} {arXiv preprint arXiv:2103.12217}\ } (\bibinfo {year} {2021})}\BibitemShut {NoStop}%
\bibitem [{\citenamefont {Aghaee}\ \emph {et~al.}(2023)\citenamefont {Aghaee}, \citenamefont {Akkala}, \citenamefont {Alam}, \citenamefont {Ali}, \citenamefont {Alcaraz~Ramirez}, \citenamefont {Andrzejczuk}, \citenamefont {Antipov}, \citenamefont {Aseev}, \citenamefont {Astafev}, \citenamefont {Bauer} \emph {et~al.}}]{aghaee2023inas}%
  \BibitemOpen
  \bibfield  {author} {\bibinfo {author} {\bibfnamefont {M.}~\bibnamefont {Aghaee}}, \bibinfo {author} {\bibfnamefont {A.}~\bibnamefont {Akkala}}, \bibinfo {author} {\bibfnamefont {Z.}~\bibnamefont {Alam}}, \bibinfo {author} {\bibfnamefont {R.}~\bibnamefont {Ali}}, \bibinfo {author} {\bibfnamefont {A.}~\bibnamefont {Alcaraz~Ramirez}}, \bibinfo {author} {\bibfnamefont {M.}~\bibnamefont {Andrzejczuk}}, \bibinfo {author} {\bibfnamefont {A.~E.}\ \bibnamefont {Antipov}}, \bibinfo {author} {\bibfnamefont {P.}~\bibnamefont {Aseev}}, \bibinfo {author} {\bibfnamefont {M.}~\bibnamefont {Astafev}}, \bibinfo {author} {\bibfnamefont {B.}~\bibnamefont {Bauer}}, \emph {et~al.},\ }\bibfield  {title} {\bibinfo {title} {Inas-al hybrid devices passing the topological gap protocol},\ }\href@noop {} {\bibfield  {journal} {\bibinfo  {journal} {Physical Review B}\ }\textbf {\bibinfo {volume} {107}},\ \bibinfo {pages} {245423} (\bibinfo {year} {2023})}\BibitemShut {NoStop}%
\bibitem [{\citenamefont {Hess}\ \emph {et~al.}(2023)\citenamefont {Hess}, \citenamefont {Legg}, \citenamefont {Loss},\ and\ \citenamefont {Klinovaja}}]{hess2023trivial}%
  \BibitemOpen
  \bibfield  {author} {\bibinfo {author} {\bibfnamefont {R.}~\bibnamefont {Hess}}, \bibinfo {author} {\bibfnamefont {H.~F.}\ \bibnamefont {Legg}}, \bibinfo {author} {\bibfnamefont {D.}~\bibnamefont {Loss}},\ and\ \bibinfo {author} {\bibfnamefont {J.}~\bibnamefont {Klinovaja}},\ }\bibfield  {title} {\bibinfo {title} {Trivial andreev band mimicking topological bulk gap reopening in the nonlocal conductance of long rashba nanowires},\ }\href@noop {} {\bibfield  {journal} {\bibinfo  {journal} {Physical Review Letters}\ }\textbf {\bibinfo {volume} {130}},\ \bibinfo {pages} {207001} (\bibinfo {year} {2023})}\BibitemShut {NoStop}%
\bibitem [{\citenamefont {Leijnse}\ and\ \citenamefont {Flensberg}(2012)}]{leijnse2012parity}%
  \BibitemOpen
  \bibfield  {author} {\bibinfo {author} {\bibfnamefont {M.}~\bibnamefont {Leijnse}}\ and\ \bibinfo {author} {\bibfnamefont {K.}~\bibnamefont {Flensberg}},\ }\bibfield  {title} {\bibinfo {title} {Parity qubits and poor man's majorana bound states in double quantum dots},\ }\href@noop {} {\bibfield  {journal} {\bibinfo  {journal} {Physical Review B—Condensed Matter and Materials Physics}\ }\textbf {\bibinfo {volume} {86}},\ \bibinfo {pages} {134528} (\bibinfo {year} {2012})}\BibitemShut {NoStop}%
\bibitem [{\citenamefont {Kraus}\ \emph {et~al.}(2013)\citenamefont {Kraus}, \citenamefont {Zoller},\ and\ \citenamefont {Baranov}}]{kraus2013braiding}%
  \BibitemOpen
  \bibfield  {author} {\bibinfo {author} {\bibfnamefont {C.~V.}\ \bibnamefont {Kraus}}, \bibinfo {author} {\bibfnamefont {P.}~\bibnamefont {Zoller}},\ and\ \bibinfo {author} {\bibfnamefont {M.~A.}\ \bibnamefont {Baranov}},\ }\bibfield  {title} {\bibinfo {title} {Braiding of atomic majorana fermions in wire networks and implementation<? format?> of the deutsch-jozsa algorithm},\ }\href@noop {} {\bibfield  {journal} {\bibinfo  {journal} {Physical Review Letters}\ }\textbf {\bibinfo {volume} {111}},\ \bibinfo {pages} {203001} (\bibinfo {year} {2013})}\BibitemShut {NoStop}%
\bibitem [{\citenamefont {Mi}\ \emph {et~al.}(2022)\citenamefont {Mi}, \citenamefont {Sonner}, \citenamefont {Niu}, \citenamefont {Lee}, \citenamefont {Foxen}, \citenamefont {Acharya}, \citenamefont {Aleiner}, \citenamefont {Andersen}, \citenamefont {Arute}, \citenamefont {Arya} \emph {et~al.}}]{mi2022noise}%
  \BibitemOpen
  \bibfield  {author} {\bibinfo {author} {\bibfnamefont {X.}~\bibnamefont {Mi}}, \bibinfo {author} {\bibfnamefont {M.}~\bibnamefont {Sonner}}, \bibinfo {author} {\bibfnamefont {M.~Y.}\ \bibnamefont {Niu}}, \bibinfo {author} {\bibfnamefont {K.~W.}\ \bibnamefont {Lee}}, \bibinfo {author} {\bibfnamefont {B.}~\bibnamefont {Foxen}}, \bibinfo {author} {\bibfnamefont {R.}~\bibnamefont {Acharya}}, \bibinfo {author} {\bibfnamefont {I.}~\bibnamefont {Aleiner}}, \bibinfo {author} {\bibfnamefont {T.~I.}\ \bibnamefont {Andersen}}, \bibinfo {author} {\bibfnamefont {F.}~\bibnamefont {Arute}}, \bibinfo {author} {\bibfnamefont {K.}~\bibnamefont {Arya}}, \emph {et~al.},\ }\bibfield  {title} {\bibinfo {title} {Noise-resilient edge modes on a chain of superconducting qubits},\ }\href@noop {} {\bibfield  {journal} {\bibinfo  {journal} {Science}\ }\textbf {\bibinfo {volume} {378}},\ \bibinfo {pages} {785} (\bibinfo {year} {2022})}\BibitemShut {NoStop}%
\bibitem [{\citenamefont {Grusdt}\ \emph {et~al.}(2016)\citenamefont {Grusdt}, \citenamefont {Yao}, \citenamefont {Abanin}, \citenamefont {Fleischhauer},\ and\ \citenamefont {Demler}}]{grusdt2016interferometric}%
  \BibitemOpen
  \bibfield  {author} {\bibinfo {author} {\bibfnamefont {F.}~\bibnamefont {Grusdt}}, \bibinfo {author} {\bibfnamefont {N.~Y.}\ \bibnamefont {Yao}}, \bibinfo {author} {\bibfnamefont {D.}~\bibnamefont {Abanin}}, \bibinfo {author} {\bibfnamefont {M.}~\bibnamefont {Fleischhauer}},\ and\ \bibinfo {author} {\bibfnamefont {E.}~\bibnamefont {Demler}},\ }\bibfield  {title} {\bibinfo {title} {Interferometric measurements of many-body topological invariants using mobile impurities},\ }\href@noop {} {\bibfield  {journal} {\bibinfo  {journal} {Nature communications}\ }\textbf {\bibinfo {volume} {7}},\ \bibinfo {pages} {11994} (\bibinfo {year} {2016})}\BibitemShut {NoStop}%
\bibitem [{\citenamefont {Grusdt}\ \emph {et~al.}(2019)\citenamefont {Grusdt}, \citenamefont {Yao},\ and\ \citenamefont {Demler}}]{grusdt2019topological}%
  \BibitemOpen
  \bibfield  {author} {\bibinfo {author} {\bibfnamefont {F.}~\bibnamefont {Grusdt}}, \bibinfo {author} {\bibfnamefont {N.~Y.}\ \bibnamefont {Yao}},\ and\ \bibinfo {author} {\bibfnamefont {E.}~\bibnamefont {Demler}},\ }\bibfield  {title} {\bibinfo {title} {Topological polarons, quasiparticle invariants, and their detection in one-dimensional symmetry-protected phases},\ }\href@noop {} {\bibfield  {journal} {\bibinfo  {journal} {Physical Review B}\ }\textbf {\bibinfo {volume} {100}},\ \bibinfo {pages} {075126} (\bibinfo {year} {2019})}\BibitemShut {NoStop}%
\bibitem [{\citenamefont {Pimenov}\ \emph {et~al.}(2021)\citenamefont {Pimenov}, \citenamefont {Camacho-Guardian}, \citenamefont {Goldman}, \citenamefont {Massignan}, \citenamefont {Bruun},\ and\ \citenamefont {Goldstein}}]{pimenov2021topological}%
  \BibitemOpen
  \bibfield  {author} {\bibinfo {author} {\bibfnamefont {D.}~\bibnamefont {Pimenov}}, \bibinfo {author} {\bibfnamefont {A.}~\bibnamefont {Camacho-Guardian}}, \bibinfo {author} {\bibfnamefont {N.}~\bibnamefont {Goldman}}, \bibinfo {author} {\bibfnamefont {P.}~\bibnamefont {Massignan}}, \bibinfo {author} {\bibfnamefont {G.~M.}\ \bibnamefont {Bruun}},\ and\ \bibinfo {author} {\bibfnamefont {M.}~\bibnamefont {Goldstein}},\ }\bibfield  {title} {\bibinfo {title} {Topological transport of mobile impurities},\ }\href@noop {} {\bibfield  {journal} {\bibinfo  {journal} {Physical Review B}\ }\textbf {\bibinfo {volume} {103}},\ \bibinfo {pages} {245106} (\bibinfo {year} {2021})}\BibitemShut {NoStop}%
\bibitem [{\citenamefont {Vashisht}\ \emph {et~al.}(2024)\citenamefont {Vashisht}, \citenamefont {Amelio}, \citenamefont {Vanderstraeten}, \citenamefont {Bruun}, \citenamefont {Diessel},\ and\ \citenamefont {Goldman}}]{vashisht2024chiral}%
  \BibitemOpen
  \bibfield  {author} {\bibinfo {author} {\bibfnamefont {A.}~\bibnamefont {Vashisht}}, \bibinfo {author} {\bibfnamefont {I.}~\bibnamefont {Amelio}}, \bibinfo {author} {\bibfnamefont {L.}~\bibnamefont {Vanderstraeten}}, \bibinfo {author} {\bibfnamefont {G.~M.}\ \bibnamefont {Bruun}}, \bibinfo {author} {\bibfnamefont {O.~K.}\ \bibnamefont {Diessel}},\ and\ \bibinfo {author} {\bibfnamefont {N.}~\bibnamefont {Goldman}},\ }\bibfield  {title} {\bibinfo {title} {Chiral polaron formation on the edge of topological quantum matter},\ }\href@noop {} {\bibfield  {journal} {\bibinfo  {journal} {arXiv preprint arXiv:2407.19093}\ } (\bibinfo {year} {2024})}\BibitemShut {NoStop}%
\bibitem [{\citenamefont {Atala}\ \emph {et~al.}(2013)\citenamefont {Atala}, \citenamefont {Aidelsburger}, \citenamefont {Barreiro}, \citenamefont {Abanin}, \citenamefont {Kitagawa}, \citenamefont {Demler},\ and\ \citenamefont {Bloch}}]{atala2013direct}%
  \BibitemOpen
  \bibfield  {author} {\bibinfo {author} {\bibfnamefont {M.}~\bibnamefont {Atala}}, \bibinfo {author} {\bibfnamefont {M.}~\bibnamefont {Aidelsburger}}, \bibinfo {author} {\bibfnamefont {J.~T.}\ \bibnamefont {Barreiro}}, \bibinfo {author} {\bibfnamefont {D.}~\bibnamefont {Abanin}}, \bibinfo {author} {\bibfnamefont {T.}~\bibnamefont {Kitagawa}}, \bibinfo {author} {\bibfnamefont {E.}~\bibnamefont {Demler}},\ and\ \bibinfo {author} {\bibfnamefont {I.}~\bibnamefont {Bloch}},\ }\bibfield  {title} {\bibinfo {title} {Direct measurement of the zak phase in topological bloch bands},\ }\href@noop {} {\bibfield  {journal} {\bibinfo  {journal} {Nature Physics}\ }\textbf {\bibinfo {volume} {9}},\ \bibinfo {pages} {795} (\bibinfo {year} {2013})}\BibitemShut {NoStop}%
\bibitem [{\citenamefont {Mohseni}\ \emph {et~al.}(2023)\citenamefont {Mohseni}, \citenamefont {Allami}, \citenamefont {Miravet}, \citenamefont {Gayowsky}, \citenamefont {Korkusinski},\ and\ \citenamefont {Hawrylak}}]{mohseni2023majorana}%
  \BibitemOpen
  \bibfield  {author} {\bibinfo {author} {\bibfnamefont {M.}~\bibnamefont {Mohseni}}, \bibinfo {author} {\bibfnamefont {H.}~\bibnamefont {Allami}}, \bibinfo {author} {\bibfnamefont {D.}~\bibnamefont {Miravet}}, \bibinfo {author} {\bibfnamefont {D.~J.}\ \bibnamefont {Gayowsky}}, \bibinfo {author} {\bibfnamefont {M.}~\bibnamefont {Korkusinski}},\ and\ \bibinfo {author} {\bibfnamefont {P.}~\bibnamefont {Hawrylak}},\ }\bibfield  {title} {\bibinfo {title} {Majorana excitons in a kitaev chain of semiconductor quantum dots in a nanowire},\ }\href@noop {} {\bibfield  {journal} {\bibinfo  {journal} {Nanomaterials}\ }\textbf {\bibinfo {volume} {13}},\ \bibinfo {pages} {2293} (\bibinfo {year} {2023})}\BibitemShut {NoStop}%
\bibitem [{\citenamefont {Qin}\ \emph {et~al.}(2019)\citenamefont {Qin}, \citenamefont {Cui},\ and\ \citenamefont {Yi}}]{qin2019polaron}%
  \BibitemOpen
  \bibfield  {author} {\bibinfo {author} {\bibfnamefont {F.}~\bibnamefont {Qin}}, \bibinfo {author} {\bibfnamefont {X.}~\bibnamefont {Cui}},\ and\ \bibinfo {author} {\bibfnamefont {W.}~\bibnamefont {Yi}},\ }\bibfield  {title} {\bibinfo {title} {Polaron in ap+ ip fermi topological superfluid},\ }\href@noop {} {\bibfield  {journal} {\bibinfo  {journal} {Physical Review A}\ }\textbf {\bibinfo {volume} {99}},\ \bibinfo {pages} {033613} (\bibinfo {year} {2019})}\BibitemShut {NoStop}%
\bibitem [{\citenamefont {Tewari}\ and\ \citenamefont {Sau}(2012)}]{PhysRevLett.109.150408}%
  \BibitemOpen
  \bibfield  {author} {\bibinfo {author} {\bibfnamefont {S.}~\bibnamefont {Tewari}}\ and\ \bibinfo {author} {\bibfnamefont {J.~D.}\ \bibnamefont {Sau}},\ }\bibfield  {title} {\bibinfo {title} {Topological invariants for spin-orbit coupled superconductor nanowires},\ }\href {https://doi.org/10.1103/PhysRevLett.109.150408} {\bibfield  {journal} {\bibinfo  {journal} {Phys. Rev. Lett.}\ }\textbf {\bibinfo {volume} {109}},\ \bibinfo {pages} {150408} (\bibinfo {year} {2012})}\BibitemShut {NoStop}%
\bibitem [{\citenamefont {Chitov}(2018)}]{chitov_sop}%
  \BibitemOpen
  \bibfield  {author} {\bibinfo {author} {\bibfnamefont {G.~Y.}\ \bibnamefont {Chitov}},\ }\bibfield  {title} {\bibinfo {title} {Local and nonlocal order parameters in the kitaev chain},\ }\href {https://doi.org/10.1103/PhysRevB.97.085131} {\bibfield  {journal} {\bibinfo  {journal} {Phys. Rev. B}\ }\textbf {\bibinfo {volume} {97}},\ \bibinfo {pages} {085131} (\bibinfo {year} {2018})}\BibitemShut {NoStop}%
\bibitem [{\citenamefont {Sidler}\ \emph {et~al.}(2017)\citenamefont {Sidler}, \citenamefont {Back}, \citenamefont {Cotlet}, \citenamefont {Srivastava}, \citenamefont {Fink}, \citenamefont {Kroner}, \citenamefont {Demler},\ and\ \citenamefont {Imamoglu}}]{sidler2017fermi}%
  \BibitemOpen
  \bibfield  {author} {\bibinfo {author} {\bibfnamefont {M.}~\bibnamefont {Sidler}}, \bibinfo {author} {\bibfnamefont {P.}~\bibnamefont {Back}}, \bibinfo {author} {\bibfnamefont {O.}~\bibnamefont {Cotlet}}, \bibinfo {author} {\bibfnamefont {A.}~\bibnamefont {Srivastava}}, \bibinfo {author} {\bibfnamefont {T.}~\bibnamefont {Fink}}, \bibinfo {author} {\bibfnamefont {M.}~\bibnamefont {Kroner}}, \bibinfo {author} {\bibfnamefont {E.}~\bibnamefont {Demler}},\ and\ \bibinfo {author} {\bibfnamefont {A.}~\bibnamefont {Imamoglu}},\ }\bibfield  {title} {\bibinfo {title} {Fermi polaron-polaritons in charge-tunable atomically thin semiconductors},\ }\href@noop {} {\bibfield  {journal} {\bibinfo  {journal} {Nature Physics}\ }\textbf {\bibinfo {volume} {13}},\ \bibinfo {pages} {255} (\bibinfo {year} {2017})}\BibitemShut {NoStop}%
\end{thebibliography}%
\end{document}